\documentclass[12pt]{iopart}
\usepackage{iopams}
\usepackage{graphicx}
\usepackage{bold-extra}
\usepackage{listings}
\usepackage{bm,color}
\usepackage[usenames,dvipsnames]{xcolor}
\usepackage{bbm}
\usepackage{subfigure}
\input{Qcircuit}

\definecolor{darkgreen}{rgb}{.1,.4,.15}
\definecolor{darkred}{rgb}{.5,.1,.2}
\definecolor{darkblue}{rgb}{.2,.3,.6}

\begin{document}

\title{Gauge subsystems, separability, and robustness in autonomous quantum memories}

\author{Gopal Sarma and Hideo Mabuchi}
\address{Edward L.~Ginzton Laboratory, Stanford University, Stanford, CA 94305, USA}

\date{\today}

\begin{abstract}
Quantum error correction provides a fertile context for exploring the interplay of feedback control, microscopic physics and noncommutative probability. In this paper we deepen our understanding of this nexus through high-level analysis of a class of quantum memory models that we have previously proposed, which implement continuous-time versions of well-known stabilizer codes in autonomous nanophotonic circuits that require no external clocking or control. We show that the presence of the gauge subsystem in the nine-qubit Bacon-Shor code allows for an optimized layout of the corresponding nanophotonic circuit that substantially ameliorates the effects of optical propagation losses, argue that code separability allows for simplified restoration feedback protocols, and propose a modified fidelity metric for quantifying the performance of realistic quantum memories. Our treatment of these topics exploits the homogeneous modeling framework of autonomous nanophotonic circuits, but the key ideas translate to the traditional setting of discrete time, measurement-based quantum error correction.
\end{abstract}

\pacs{03.67.Pp,42.50.Ex,89.20.Kk}


\maketitle

\noindent Quantum feedback control provides a systems engineering perspective on the analysis and design of quantum memories, complementing alternative ideas extending from theoretical physics. Whereas the latter approach has emphasized the connections of quantum error correction and quantum computation to many body physics~\cite{Kitaev, Dennis}, the quantum control community has viewed decoherence suppression as a problem that should ideally be formulated as a non-commutative generalization of classical stochastic and hybrid control theory. In a sense, we can view this research program as a 1950's/60's-era agenda translated forward a half century. That is, given the novel resources of photonics, quantum electronics, and spintronics, we can ask ourselves what the basic constituent components might be of a post-classical information processing machine, and how the exchange of signals among such components could be used to realize desired functionality~\cite{Ahn,BvHJ,Mabu09a}.

Within this setting we have recently investigated nanophotonic circuit models for quantum memories that autonomously implement well-known stabilizer codes. In these models, each physical qubit is strongly coupled to an optical or microwave resonator, and the resonators are coupled by waveguides to form a coherent feedback network (photonic circuit). When the circuit is powered by appropriate stationary laser inputs, the resulting continuous-time dynamics effect error detection and correction without any additional external clocking or control signals. In such models, the classical signal processing apparatus assumed in standard treatments of quantum error correction is replaced by a small number of controller qubits, making it possible to derive a master equation for the closed-loop behavior that can be modified straightforwardly to incorporate a wide range of realistic decoherence mechanisms~\cite{Kerc10,Kerc11,Sarma12a}.

The operational principles of these quantum memory models, as well as the methods used to derive their equations of motion, exemplify a quantum-optical generalization of conventional electric circuit theory in which guided electromagnetic fields play the role of signals and qubit-resonators serve as input/output devices that process them coherently. We have previously shown how the the Gough-James quantum network algebra~\cite{GJ09a,GJ09b,Goug10} can be utilized together with a recent limit theorem for quantum stochastic differential equations (QSDEs)~\cite{Bout08} to facilitate the derivation of an intuitive master equation for a given quantum memory model from an explicit construction of the underlying nanophotonic circuit, in a manner inspired by schematic capture methods of contemporary electrical engineering~\cite{Sarma12a,Teza12}. Here, in order to focus on higher architectural principles, we will skip over such details and jump directly to considering a class of master equations that arises from the general approach of embedding continuous-time relaxations of stabilizer quantum error correcting codes in the autonomous dynamics of a coherent feedback network.

We model an autonomous quantum memory using a Markovian master equation~\cite{GZ04},
\begin{equation}
\dot{\rho} = -i[H, \rho] + \sum_{i}\left\{L_{i}\rho L_{i}^{\dagger} - \frac{1}{2}L_{i}^{\dagger}L_{i}\rho - \frac{1}{2}\rho L_{i}^{\dagger}L_{i}\right\},
\end{equation}
where $H$ is a Hamiltonian for the internal dynamics of the memory and controller qubits, and the Lindblad operators $L_{i}$ describe couplings to reservoir modes that mediate decoherence (including memory errors) and entropy removal (via spontaneous emission-type processes). At this level of description, the dynamics of the electromagnetic field modes within the waveguides and resonators of the photonic circuit have been adiabatically eliminated. Details of the derivation of such a master equation for the three-qubit code can be found in~\cite{Kerc10}; here in considering arbitrary stabilizer codes we assume that the register qubit model~\cite{Kerc09} can be generalized to incorporate atomic level structures whose Raman resonance conditions effectively implement an \textbf{\texttt{AND}} operation on as many feedback signals as the code requires. We assume a feedback controller construction based on nanophotonic relays~\cite{Mabu09} as in prior work.

Consider a stabilizer quantum error correcting code~\cite{Gottesman, MikeIke} that encodes one qubit of information in $Q$ register qubits with $N$ stabilizer generators. We refer to the complete set of stabilizer generators as $\mathcal{S} = \{M_{n}\}_{n=1}^{N}$. Recall that a stabilizer code redundantly encodes information in the joint state of the register qubits, that each stabilizer generator is a joint observable of the register qubits with eigenvalues $\pm 1$, and that a measurement of the full set of stabilizer generators values (the error {\it syndrome}) suffices to localize any correctable error but yields no information on the encoded qubit. To explicitly construct the master equation for an autonomous quantum memory based on $\mathcal{S}$ we use the following procedure. For each stabilizer generator $M_n$ include two Lindblad operators of the form
\begin{eqnarray}
L_{2n-1} &=& \alpha\left(\sigma_{+}^{R_{n}}(I + M_{n}) - \Pi_{g}^{R_{n}}(I - M_{n})\right),\\
L_{2n} &=& \alpha\left(\sigma_{-}^{R_{n}}(I - M_{n}) + \Pi_{h}^{R_{n}}(I + M_{n})\right),
\end{eqnarray}
where $\sigma_{\pm}^{R_{n}}$, $\Pi_{g}^{R_{n}}$ and $\Pi_{h}^{R_{n}}$ are (qubit-like) raising/lowering operators and projectors onto $g$ and $h$ basis states for the $n^{\rm th}$ relay of the feedback controller, and $\alpha$ parameterizes the amplitudes of the electromagnetic probe fields, which in turn determines the syndrome measurement rate. If the code is {\em separable}, in the sense that disjoint sets of stabilizer generators mediate $X$ and $Z$ error syndrome extraction, the quantum memory Hamiltonian can be written in the form
\begin{equation}
H = \sum_{n=1}^Q \Omega\left(X_{n}\cdot \mathcal{F}_{\mathcal{S}_{X}}\left[X_{n}\right] + Z_{n}\cdot \mathcal{F}_{\mathcal{S}_{Z}}\left[Z_{n}\right] \right).
\end{equation}
If the code is not separable then
\begin{equation}
H = \sum_{n=1}^Q \Omega\left(X_{n}\cdot \mathcal{F}_{\mathcal{S}}\left[X_{n}\right] + Y_{n}\cdot \mathcal{F}_{\mathcal{S}}\left[Y_{n}\right] + Z_{n}\cdot \mathcal{F}_{\mathcal{S}}\left[Z_{n}\right]\right),
\end{equation}
where the function $\mathcal{F}_{\mathcal{S}}\left[E_{n}\right]$ maps a single-qubit error to a projector onto a state of the controller relays. The action of $\mathcal{F}$ straightforwardly represents the lookup table of correctable errors and corresponding syndromes. For example, suppose there are six stabilizer generators and that they take the values 1 -1 -1 1 -1 1 for a register state with a bit-flip error on the fourth qubit. Then $\mathcal{F}_{\mathcal{S}}\left[X_{4}\right] = \Pi_{h}^{R_{1}} \Pi_{g}^{R_{2}} \Pi_{g}^{R_{3}} \Pi_{h}^{R_{4}} \Pi_{g}^{R_{5}} \Pi_{h}^{R_{6}}$ (the $h$ state of relay $n$ is associated with the $+1$ value of $M_n$, and the $g$ state with value $-1$). When the code is separable, we use $\mathcal{S}_{X}$ and $\mathcal{S}_{Z}$ to denote disjoint subsets of $\mathcal{S}$. In both of the above Hamiltonian expressions, $\Omega$ parameterizes the strength of the feedback fields.

To illustrate our construction we first consider the seven-qubit code~\cite{Gottesman, MikeIke}, which is separable. The following lookup table gives the corresponding error syndromes for single-qubit $X$, $Z$, and $Y$ errors with $\blacksquare$ corresponding to a $+1$ value for a given stabilizer generator and $\bigcirc$ to $-1$:

\begin{center}
\small
\begin{tabular*}{0.8\textwidth}{@{\extracolsep{\fill}} |c | c c c c c c |}
\multicolumn{7}{c}{\textbf{Error syndromes for 7-qubit code}}\\
\hline
\textbf{Error} & \textbf{M1} & \textbf{M2} & \textbf{M3} & \textbf{M4} & \textbf{M5} & \textbf{M6}\\
\hline
X1 & $\blacksquare$ & $\blacksquare$ & $\blacksquare$ & $\bigcirc$ & $\bigcirc$ & $\bigcirc$\\
X2 & $\blacksquare$ & $\blacksquare$ & $\blacksquare$ & $\bigcirc$ & $\bigcirc$ & $\blacksquare$\\
X3 & $\blacksquare$ & $\blacksquare$ & $\blacksquare$ & $\bigcirc$ & $\blacksquare$ & $\bigcirc$\\
X4 & $\blacksquare$ & $\blacksquare$ & $\blacksquare$ & $\bigcirc$ & $\blacksquare$ & $\blacksquare$ \\
X5 & $\blacksquare$ & $\blacksquare$ & $\blacksquare$ & $\blacksquare$ & $\bigcirc$ & $\bigcirc$\\
X6 & $\blacksquare$ & $\blacksquare$ & $\blacksquare$ & $\blacksquare$ & $\bigcirc$ & $\blacksquare$ \\
X7 & $\blacksquare$ & $\blacksquare$ & $\blacksquare$ & $\blacksquare$ & $\blacksquare$ & $\bigcirc$\\
\hline
Z1 & $\bigcirc$ & $\bigcirc$ & $\bigcirc$ & $\blacksquare$ & $\blacksquare$ & $\blacksquare$\\
Z2 & $\bigcirc$ & $\bigcirc$ & $\blacksquare$ & $\blacksquare$ & $\blacksquare$ & $\blacksquare$\\
Z3 & $\bigcirc$ & $\blacksquare$ & $\bigcirc$ & $\blacksquare$ & $\blacksquare$ & $\blacksquare$\\
Z4 & $\bigcirc$ & $\blacksquare$ & $\blacksquare$ & $\blacksquare$ & $\blacksquare$ & $\blacksquare$\\
Z5 & $\blacksquare$ & $\bigcirc$ & $\bigcirc$ & $\blacksquare$ & $\blacksquare$ & $\blacksquare$\\
Z6 & $\blacksquare$ & $\bigcirc$ & $\blacksquare$ & $\blacksquare$ & $\blacksquare$ & $\blacksquare$\\
Z7 & $\blacksquare$ & $\blacksquare$ & $\bigcirc$ & $\blacksquare$ & $\blacksquare$ & $\blacksquare$\\
\hline
Y1 & $\bigcirc$ & $\bigcirc$ & $\bigcirc$ & $\bigcirc$ & $\bigcirc$ & $\bigcirc$\\
Y2 & $\bigcirc$ & $\bigcirc$ & $\blacksquare$ & $\bigcirc$ & $\bigcirc$ & $\blacksquare$\\
Y3 & $\bigcirc$ & $\blacksquare$ & $\bigcirc$ & $\bigcirc$ & $\blacksquare$ & $\bigcirc$\\
Y4 & $\bigcirc$ & $\blacksquare$ & $\blacksquare$ & $\bigcirc$ & $\blacksquare$ & $\blacksquare$\\
Y5 & $\blacksquare$ & $\bigcirc$ & $\bigcirc$ & $\blacksquare$ & $\bigcirc$ & $\bigcirc$\\
Y6 & $\blacksquare$ & $\bigcirc$ & $\blacksquare$ & $\blacksquare$ & $\bigcirc$ & $\blacksquare$\\
Y7 & $\blacksquare$ & $\blacksquare$ & $\bigcirc$ & $\blacksquare$ & $\blacksquare$ & $\bigcirc$\\
\hline
 \end{tabular*}
\medskip
\end{center}

\noindent It is easily seen that the error syndrome for $Y_n$ is simply the logical \textbf{\texttt{OR}} of the syndromes for $X_n$ and $Z_n$, which combined with the algebraic relation $Y_n\propto X_nZ_n$ makes it possible to design a feedback network that independently detects and corrects $X$ and $Z$ errors---when a $Y_n$ error occurs it can be diagnosed and treated as the occurrence of both an $X_n$ and a $Z_n$ error.

Applying the construction described above for the separable case, we arrive at the following master equation for a coherent feedback implementation of the seven-qubit code:
\begin{eqnarray}
H = & \Omega\biggl( X_1\Pi_{g}^{R_4}\Pi_{g}^{R_5}\Pi_{g}^{R_6} + X_2\Pi_{g}^{R_4}\Pi_{g}^{R_5}\Pi_{h}^{R_6} + X_3\Pi_{g}^{R_4}\Pi_{h}^{R_5}\Pi_{g}^{R_6} +\nonumber\\
&X_4\Pi_{g}^{R_4}\Pi_{h}^{R_5}\Pi_{h}^{R_6} + X_5\Pi_{h}^{R_4}\Pi_{g}^{R_5}\Pi_{g}^{R_6} + X_6\Pi_{h}^{R_4}\Pi_{g}^{R_5}\Pi_{h}^{R_6} +\nonumber\\
&X_7\Pi_{h}^{R_4}\Pi_{h}^{R_5}\Pi_{g}^{R_6} + Z_1\Pi_{g}^{R_1}\Pi_{g}^{R_2}\Pi_{g}^{R_3} + Z_2\Pi_{g}^{R_1}\Pi_{g}^{R_2}\Pi_{h}^{R_3} +\nonumber\\
&Z_3\Pi_{g}^{R_1}\Pi_{h}^{R_2}\Pi_{g}^{R_3} + Z_4\Pi_{g}^{R_1}\Pi_{h}^{R_2}\Pi_{h}^{R_3} + Z_5\Pi_{h}^{R_1}\Pi_{g}^{R_2}\Pi_{g}^{R_3} +\nonumber\\
&Z_6\Pi_{h}^{R_1}\Pi_{g}^{R_2}\Pi_{h}^{R_3} + Z_7\Pi_{h}^{R_1}\Pi_{h}^{R_2}\Pi_{g}^{R_3}\biggr) ,
\end{eqnarray}
\begin{eqnarray}
L_{1} &=& \alpha\left(\sigma_{+}^{R_{1}}(I + X_1X_2X_3X_4) - \Pi_{g}^{R_{1}}(I - X_1X_2X_3X_4)\right),\nonumber\\
L_{2} &=& \alpha\left(\sigma_{-}^{R_{1}}(I - X_1X_2X_3X_4) + \Pi_{h}^{R_{1}}(I + X_1X_2X_3X_4)\right),\nonumber\\
L_{3} &=& \alpha\left(\sigma_{+}^{R_{2}}(I + X_1X_2X_5X_6) - \Pi_{g}^{R_{2}}(I - X_1X_2X_5X_6)\right),\nonumber\\
L_{4} &=& \alpha\left(\sigma_{-}^{R_{2}}(I - X_1X_2X_5X_6) + \Pi_{h}^{R_{2}}(I + X_1X_2X_5X_6)\right),\nonumber\\
L_{5} &=& \alpha\left(\sigma_{+}^{R_{3}}(I + X_1X_3X_5X_7) - \Pi_{g}^{R_{3}}(I - X_1X_3X_5X_7)\right),\nonumber\\
L_{6} &=& \alpha\left(\sigma_{-}^{R_{3}}(I - X_1X_3X_5X_7) + \Pi_{h}^{R_{3}}(I + X_1X_3X_5X_7)\right),\nonumber\\
L_{7} &=& \alpha\left(\sigma_{+}^{R_{4}}(I + Z_1Z_2Z_3Z_4) - \Pi_{g}^{R_{4}}(I - Z_1Z_2Z_3Z_4)\right),\nonumber\\
L_{8} &=& \alpha\left(\sigma_{-}^{R_{4}}(I - Z_1Z_2Z_3Z_4) + \Pi_{h}^{R_{4}}(I + Z_1Z_2Z_3Z_4)\right),\nonumber\\
L_{9} &=& \alpha\left(\sigma_{+}^{R_{5}}(I + Z_1Z_2Z_5Z_6) - \Pi_{g}^{R_{5}}(I - Z_1Z_2Z_5Z_6)\right),\nonumber\\
L_{10} &=& \alpha\left(\sigma_{-}^{R_{5}}(I - Z_1Z_2Z_5Z_6) + \Pi_{h}^{R_{5}}(I + Z_1Z_2Z_5Z_6)\right),\nonumber\\
L_{11} &=& \alpha\left(\sigma_{+}^{R_{6}}(I + Z_1Z_3Z_5Z_7) - \Pi_{g}^{R_{6}}(I - Z_1Z_3Z_5Z_7)\right),\nonumber\\
L_{12} &=& \alpha\left(\sigma_{-}^{R_{6}}(I - Z_1Z_3Z_5Z_7) + \Pi_{h}^{R_{6}}(I + Z_1Z_3Z_5Z_7)\right).
\medskip
\end{eqnarray}
Any desired Markovian error model can be incorporated via additional Lindblad terms, {\it e.g.}, for bit-flip errors,
\begin{equation}
L_{12+n}=\sqrt{\Gamma}X_n,\quad n\in 1\ldots Q,
\end{equation}
or for spontaneous emission-type decoherence,
\begin{equation}
L_{12+n}=\sqrt{\Gamma}(X_n - iY_n),\quad n\in 1\ldots Q.
\end{equation}
In either case, the parameter $\Gamma$ adjusts the decoherence rate. Noise processes acting on the controller degrees of freedom can be included in the analogous fashion.

If we next consider the five-qubit code, the smallest quantum error correcting code capable of protecting a single encoded qubit against arbitrary single-qubit errors~\cite{Gottesman, MikeIke}, the following lookup table gives the error syndromes for $X$, $Z$ and $Y$ errors acting on the register qubits:

\begin{center}
\small
\begin{tabular*}{0.7\textwidth}{@{\extracolsep{\fill}} |c | c c c l |}
\multicolumn{5}{c}{\textbf{Error syndromes for 5-qubit code}}\\
\hline
\textbf{Error} & \textbf{M1} & \textbf{M2} & \textbf{M3} & \textbf{M4}\\
\hline
X1 & $\blacksquare$ & $\bigcirc$ & $\blacksquare$ & $\blacksquare$\\
X2 & $\bigcirc$ & $\blacksquare$ & $\bigcirc$ & $\blacksquare$\\
X3 & $\blacksquare$ & $\bigcirc$ & $\blacksquare$ & $\bigcirc$\\
X4 & $\blacksquare$ & $\blacksquare$ & $\bigcirc$ & $\blacksquare$\\
X5 & $\bigcirc$ & $\blacksquare$ & $\blacksquare$ & $\bigcirc$\\
\hline
Z1 & $\blacksquare$ & $\blacksquare$ & $\bigcirc$ & $\blacksquare$\\
Z2 & $\blacksquare$ & $\blacksquare$ & $\bigcirc$ & $\bigcirc$\\
Z3 & $\bigcirc$ & $\blacksquare$ & $\blacksquare$ & $\blacksquare$\\
Z4 & $\bigcirc$ & $\bigcirc$ & $\blacksquare$ & $\blacksquare$\\
Z5 & $\blacksquare$ & $\bigcirc$ & $\bigcirc$ & $\blacksquare$\\
\hline
Y1 & $\blacksquare$ & $\bigcirc$ & $\bigcirc$ & $\bigcirc$\\
Y2 & $\bigcirc$ & $\blacksquare$ & $\bigcirc$ & $\bigcirc$\\
Y3 & $\bigcirc$ & $\bigcirc$ & $\blacksquare$ & $\bigcirc$\\
Y4 & $\bigcirc$ & $\bigcirc$ & $\bigcirc$ & $\blacksquare$\\
Y5 & $\bigcirc$ & $\bigcirc$ & $\bigcirc$ & $\bigcirc$\\
\hline
 \end{tabular*}
\medskip
\end{center}

\noindent It appears by inspection that the syndromes of $X_n$, $Z_n$ and $Y_n$ are not simply related, so evidently we must implement a feedback controller with sufficient logic to react conditionally to each of the fifteen four-bit syndromes. For the separable seven-qubit code we have seen that only two independent sets of seven three-bit syndromes need to be interpreted, suggesting that the feedback control sub-circuit in an autonomous quantum memory based on the seven-qubit code could be substantially simpler than for the five-qubit code, {\em even though it involves two additional stabilizer generators}. Given that we have in mind a homogeneous implementation paradigm, in which the quantum error-correcting controller is constructed from components that are very similar in nature to those of the codeword register~\cite{Kerc10}, this comparison suggests a general advantage of separable codes in terms of implementation circuit complexity and concomitant physical resource requirements.

Applying the construction described above for a non-separable code, we arrive at the following master equation for a coherent feedback implementation of the five-qubit code:
\begin{eqnarray}
H =& \Omega\biggl( X_1\Pi_{h}^{R_1}\Pi_{g}^{R_2}\Pi_{h}^{R_3}\Pi_{h}^{R_4}
+X_2\Pi_{g}^{R_1}\Pi_{h}^{R_2}\Pi_{g}^{R_3}\Pi_{h}^{R_4}
+X_3\Pi_{h}^{R_1}\Pi_{g}^{R_2}\Pi_{h}^{R_3}\Pi_{g}^{R_4} +\nonumber\\
&X_4\Pi_{h}^{R_1}\Pi_{h}^{R_2}\Pi_{g}^{R_3}\Pi_{h}^{R_4}
+X_5\Pi_{g}^{R_1}\Pi_{h}^{R_2}\Pi_{h}^{R_3}\Pi_{g}^{R_4} +\nonumber\\
&Z_1\Pi_{h}^{R_1}\Pi_{h}^{R_2}\Pi_{g}^{R_3}\Pi_{g}^{R_4}
+Z_2\Pi_{h}^{R_1}\Pi_{h}^{R_2}\Pi_{h}^{R_3}\Pi_{g}^{R_4}
+Z_3\Pi_{g}^{R_1}\Pi_{h}^{R_2}\Pi_{h}^{R_3}\Pi_{h}^{R_4} +\nonumber\\
&Z_4\Pi_{g}^{R_1}\Pi_{g}^{R_2}\Pi_{h}^{R_3}\Pi_{h}^{R_4}
+Z_5\Pi_{h}^{R_1}\Pi_{g}^{R_2}\Pi_{g}^{R_3}\Pi_{h}^{R_4} +\nonumber\\
&Y_1\Pi_{h}^{R_1}\Pi_{g}^{R_2}\Pi_{g}^{R_3}\Pi_{g}^{R_4}
+Y_2\Pi_{g}^{R_1}\Pi_{h}^{R_2}\Pi_{g}^{R_3}\Pi_{g}^{R_4}
+Y_3\Pi_{g}^{R_1}\Pi_{g}^{R_2}\Pi_{h}^{R_3}\Pi_{g}^{R_4} +\nonumber\\
&Y_4\Pi_{g}^{R_1}\Pi_{g}^{R_2}\Pi_{g}^{R_3}\Pi_{h}^{R_4}
+Y_5\Pi_{g}^{R_1}\Pi_{g}^{R_2}\Pi_{g}^{R_3}\Pi_{g}^{R_4}),
\end{eqnarray}
\begin{eqnarray}
L_{1} &=& \alpha\left(\sigma_{+}^{R_{1}}(I + Z_2X_3X_4Z_5) - \Pi_{g}^{R_{1}}(I - Z_2X_3X_4Z_5)\right),\nonumber\\
L_{2} &=& \alpha\left(\sigma_{-}^{R_{1}}(I - Z_2X_3X_4Z_5) + \Pi_{h}^{R_{1}}(I + Z_2X_3X_4Z_5)\right),\nonumber\\
L_{3} &=& \alpha\left(\sigma_{+}^{R_{2}}(I + Z_1Z_3X_4X_5) - \Pi_{g}^{R_{2}}(I - Z_1Z_3X_4X_5)\right),\nonumber\\
L_{4} &=& \alpha\left(\sigma_{-}^{R_{2}}(I - Z_1Z_3X_4X_5) + \Pi_{h}^{R_{2}}(I + Z_1Z_3X_4X_5)\right),\nonumber\\
L_{5} &=& \alpha\left(\sigma_{+}^{R_{3}}(I + X_1Z_2Z_4X_5) - \Pi_{g}^{R_{3}}(I - X_1Z_2Z_4X_5)\right),\nonumber\\
L_{6} &=& \alpha\left(\sigma_{-}^{R_{3}}(I - X_1Z_2Z_4X_5) + \Pi_{h}^{R_{3}}(I + X_1Z_2Z_4X_5)\right),\nonumber\\
L_{7} &=& \alpha\left(\sigma_{+}^{R_{4}}(I + X_1X_2Z_3Z_5) - \Pi_{g}^{R_{4}}(I - X_1X_2Z_3Z_5)\right),\nonumber\\
L_{8} &=& \alpha\left(\sigma_{-}^{R_{4}}(I - X_1X_2Z_3Z_5) + \Pi_{h}^{R_{4}}(I + X_1X_2Z_3Z_5)\right).
\end{eqnarray}
Again, decoherence processes acting on the register qubits and/or the controller degrees of freedom can be incorporated using additional Lindblad terms.

The modeling approach that we have described also admits a straightforward extension to incorporate the effects of optical propagation losses in the waveguides that connect components within the nanophotonic circuit~\cite{Sarma12a}, which lead to important considerations of optimal circuit layout that will be discussed below. As an illustrative example we first consider the probe subnetwork that extracts the error syndrome for the first stabilizer generator in the five-qubit code, $Z_2X_3X_4Z_5$. This measurement is implemented by sequentially interrogating the second, third, fourth and fifth register qubit-resonator components with a coherent laser field. In an idealized model with no optical propagation losses for the probe field, the net effect of these couplings is represented completely by the pair of coupling terms $L_1$ and $L_2$ given in the above section. In a more realistic model that accounts for optical waveguide losses, however, information leaks out into the environment as the probe field propagates between components in the photonic circuit. It can be shown~\cite{Sarma12a} that the resulting decoherence processes are described by additional coupling terms which amount to errors, some of which are correlated errors of multiple qubits that the five-bit code is not designed to correct:
\begin{eqnarray}
L_n = \alpha\theta Z_5, & L_{n+1} = \alpha\theta X_4Z_5,\nonumber\\
L_{n+2} = \alpha\theta X_3X_4Z_5,\quad & L_{n+3} = \alpha\theta Z_2X_3X_4Z_5.\label{eq:corr5}
\end{eqnarray}
Here $\theta$ parameterizes the probe field loss per waveguide segment. Similar sets of additional errors would arise from losses along the probe field paths associated with each of the other three stabilizer generators.

While our analysis leading to Eqs.~(\ref{eq:corr5}) has been grounded in the specialized modeling framework of autonomous nanophotonic circuits, our findings generally parallel known results from discrete-time, measurement-based implementation scenarios. We obtained correlated error processes by considering probe field propagation losses in a coherent feedback network, but analogous difficulties would result from any syndrome extraction mechanism in which the ancillary qubits used to accumulate the stabilizer generator values are subject to decoherence. For example, an idealized continuous-time measurement of the parity of a pair of register qubits $Q_1$, $Q_2$ in our framework~\cite{Kerc09,Kerc10} can be thought of as corresponding to the standard quantum computational circuit diagram on the left, below:

\begin{centering}
\vbox{\smallskip}
\Qcircuit @C=1em @R=.7em {
&& A && \gate{H} & \ctrl{1} & \ctrl{2} &  \gate{H} & \qw & \meter & & &
A && \gate{H} & \ctrl{1} & \gate{E} & \ctrl{2} &  \gate{H} & \qw & \meter \\
&& Q_1 && \qw & \gate{Z} & \qw &  \qw &  \qw &  \qw & & &
Q_1 && \qw & \gate{Z} & \qw &  \qw &  \qw &  \qw &  \qw \\
&& Q_2 && \qw & \qw & \gate{Z} &  \qw &  \qw &  \qw & & &
Q_2 && \qw & \qw & \qw & \gate{Z} &  \qw &  \qw &  \qw
}
\vbox{\bigskip}
\end{centering}

\noindent The propagation losses we have considered essentially correspond to an error process acting on the ancillary qubit $A$, which takes place between the controlled-Z gates as indicated by the $E$ gate in the above-right diagram. Of course, ancilla decoherence in quantum error correction has been treated in detail in the literature on fault-tolerant computing, as described for example in~\cite{Preskill97,Cody11}. In our context it is natural to assume that the dominant type of error process acting on the syndrome probe fields is optical loss, and in what follows we will show that it is possible to improve the robustness of the type of autonomous quantum memory we consider simply by optimizing the circuit layout.

There turns out to be an interesting connection between circuit layout and robustness to propagation losses in our approach, for autonomous quantum memories based on subsystem codes. Also known as operator quantum error correcting codes, subsystem codes are generalizations of decoherence free subspaces, noiseless subsystems, and quantum error correcting codes, which have gained popularity in recent years because of the large class of encoded logical operators these codes induce, which allows for simplified error recovery~\cite{Bacon2006, Poulin2005}. In~\cite{Kerc11} we considered a nanophotonic circuit for implementing a continuous-time version of the Bacon-Shor code, and noted that it is possible to reduce the circuit complexity by taking advantage of the subsystem flexibility in choosing register-qubit operations for corrective feedback. Here we further note that the subsystem structure also presents key advantages for syndrome extraction, as (following a fundamental insight discussed by Aliferis and Cross~\cite{AlifCross}) we can route the probe fields along paths that push the extra errors induced by optical propagation losses onto the unimportant gauge qubit degrees of freedom. Consequently, the correlated errors described above for the five-qubit code and in~\cite{Sarma12a} for the nine-qubit code are no longer present and the storage fidelity of the encoded qubit is substantially increased. We will quantify the performance improvement using numerical simulations, below.

\begin{figure}[ht]
\centering
\includegraphics[width=0.48\textwidth]{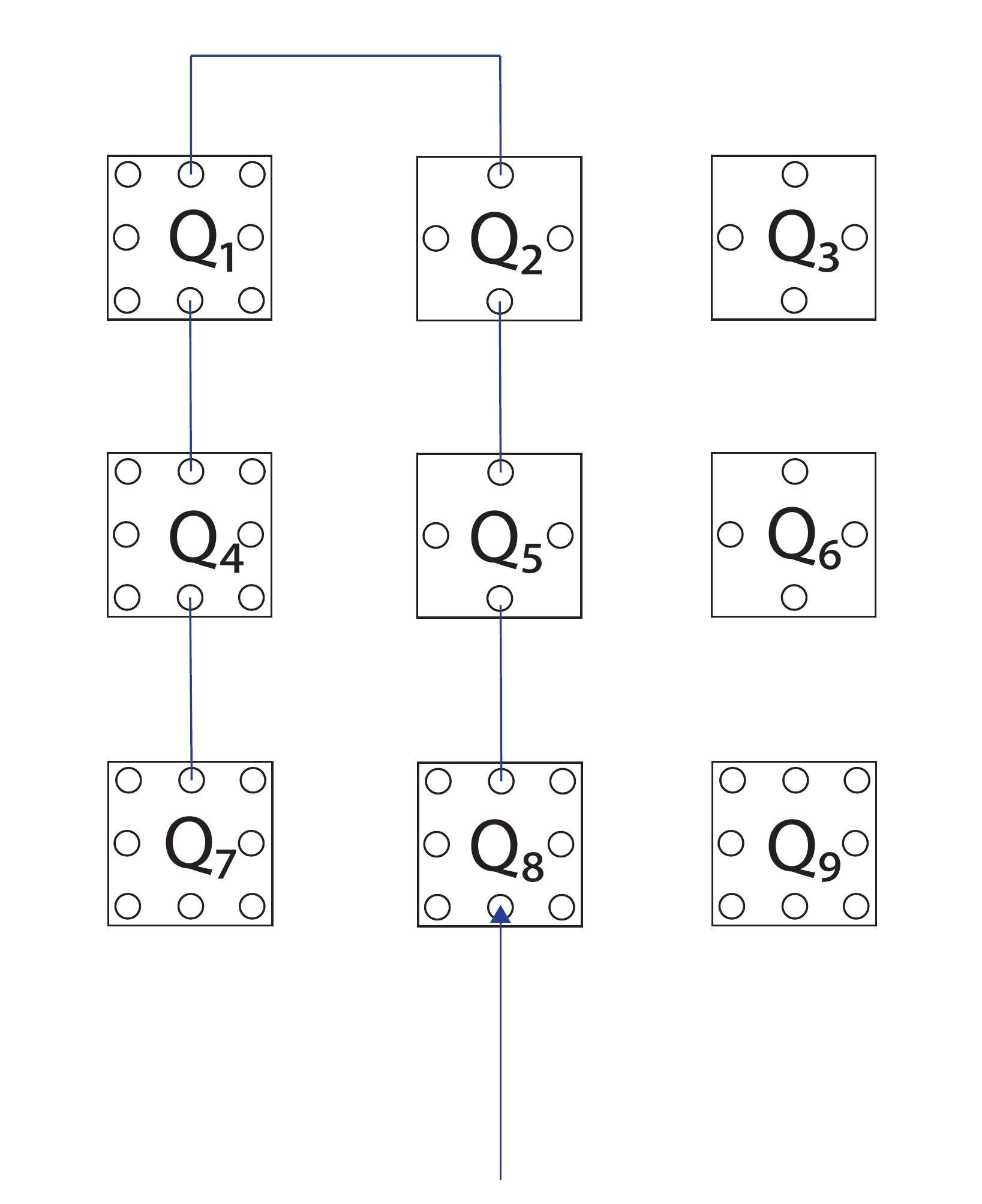}
\hspace{0.5cm}
\includegraphics[width=0.47\textwidth]{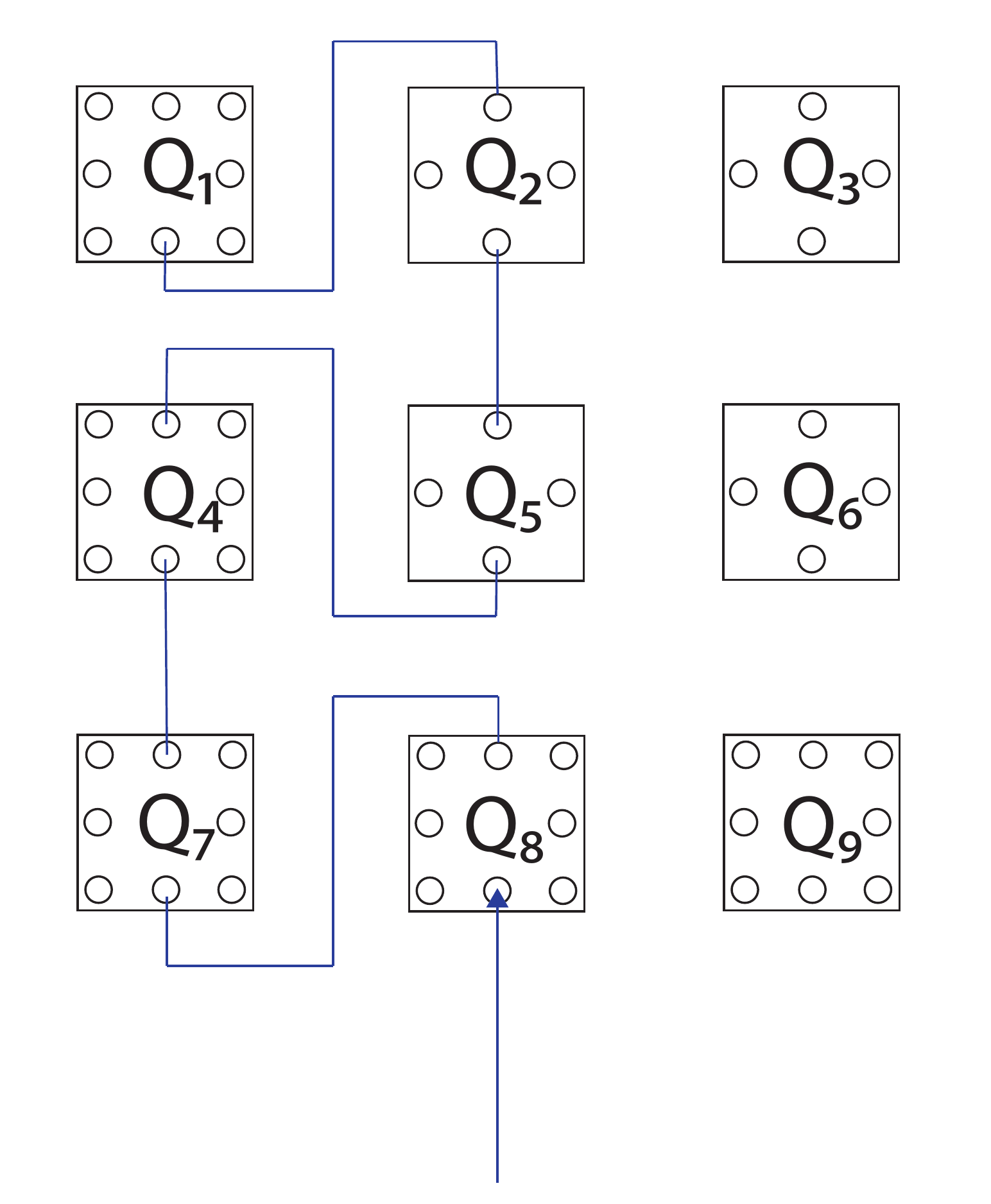}
\caption{Left: Standard probe network for Bacon-Shor nine-qubit code. Right: Subsystem routing for optimal loss protection.}
\label{fig:BS}
\end{figure}

\begin{figure}[h]
\begin{center}
\includegraphics[width=0.65\textwidth]{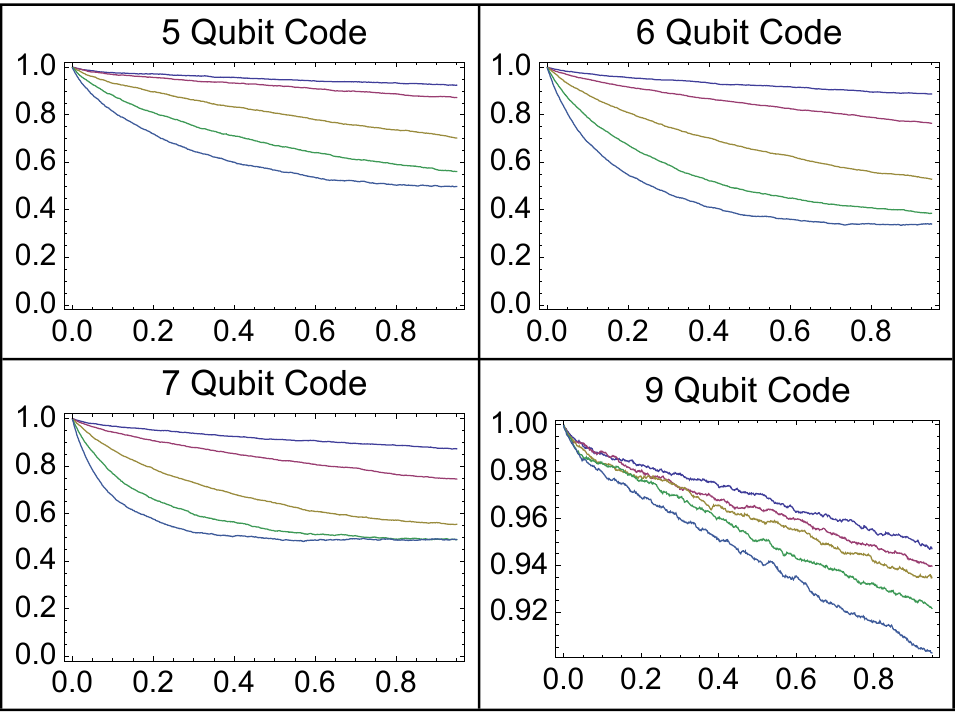}
\caption{\label{fig:lossy-qec} Decay of fidelity $\langle \Psi_{0} | \rho_{t} | \Psi_{0}\rangle$ for the five-, six-, seven- and nine-qubit quantum error correcting codes with loss parameters $\theta=\{0,1,2.5,5,7.5,10\}\pi/1000$ (top to bottom curves). For consistency with~\cite{Kerc10} the feedback strength $\Omega = \frac{|\beta|^{2} \gamma}{2\Delta}$ is set to a constant value of $200$ in each case. Note the substantial difference in the scale of the Y-axis for the bottom right corner plot---this is for the nine-qubit Bacon-Shor code with the optimally permuted probe network.}
\end{center}
\end{figure}

To illustrate the robust routing strategy, we consider the Bacon-Shor six-body stabilizer generator $Z_8Z_7Z_5Z_4Z_2Z_1$, which is one of two such $Z$-string operators for the bit-flip subnetwork~\cite{Kerc11,Sarma12a,AlifCross}. In our nanophotonic circuit model, optical propagation losses experienced by the probe field used to monitor this stabilizer generator will give rise to the following coupling terms, four of which are network-induced errors that the code is not designed to correct (while $L_n$ is a correctable error and $L_{n+5}$ merely dephases the stabilizer generator eigenstates, which does not compromise the quantum memory):
\begin{eqnarray}
L_{n} = \alpha\theta Z_8, & L_{n+1} = \alpha\theta Z_5Z_8,\nonumber\\
L_{n+2} = \alpha\theta Z_2Z_5Z_8, & L_{n+3} = \alpha\theta Z_1Z_2Z_5Z_8,\nonumber\\
L_{n+4} = \alpha\theta Z_4Z_1Z_2Z_5Z_8, \quad & L_{n+5} = \alpha\theta Z_7Z_4Z_1Z_2Z_5Z_8.\label{eq:BSsce}
\end{eqnarray}
In deriving the above set of coupling terms we have assumed a geometrically simple routing of the probe beam, assuming the register qubit-resonator components are arranged in a $3\times 3$ grid (which allows us to simplify the feedback signal routing as discussed in~\cite{Kerc11}), as shown in the left-hand schematic of Fig.~\ref{fig:BS}. The scattering order is $Z_8\rightarrow Z_5\rightarrow Z_2\rightarrow Z_1\rightarrow Z_4\rightarrow Z_7$, which bears a clear relation to the correlated error terms shown in Eqs.~(\ref{eq:BSsce}). However, we can measure the same stabilizer generator by routing the probe beam to the components in a different order as shown in the right-hand schematic of Fig.~\ref{fig:BS}, $Z_8\rightarrow Z_7\rightarrow Z_4\rightarrow Z_5\rightarrow Z_2\rightarrow Z_1$. In an idealized model with no propagation losses the scattering order should make no difference since the single-qubit Pauli operators commute. However, with propagation losses the second routing scheme leads to the following coupling terms in place of those of Eqs.~(\ref{eq:BSsce}):
\begin{eqnarray}
L_{n+1} = \alpha\theta Z_8, & L_{n+2} = \alpha\theta Z_7Z_8,\nonumber\\
L_{n+3} = \alpha\theta Z_4Z_7Z_8, & L_{n+4} = \alpha\theta Z_5Z_4Z_7Z_8,\nonumber\\
L_{n+5} = \alpha\theta Z_2Z_5Z_4Z_7Z_8, \quad & L_{n+6} = \alpha\theta Z_1Z_2Z_5Z_4Z_7Z_8.\label{eq:BSoce}
\end{eqnarray}
In the subsystem structure of the Bacon-Shor code, the operators $Z_1Z_2$, $Z_2Z_3$, $Z_4Z_5$, $Z_5Z_6$, $Z_7Z_8$, and $Z_8Z_9$ are all logical operators on the unimportant gauge-qubit degrees of freedom. Consequently, each of the coupling terms in Eqs.~(\ref{eq:BSoce}) corresponds to either a single logical qubit error or the product of a logical qubit error and one or more gauge qubit errors. Since the single qubit errors taking place on the logical space are protected by the network, this implementation of the probe mechanism is substantially more tolerant of propagation losses, as we illustrate using numerical simulations in Fig.~\ref{fig:lossy-qec}.

\begin{figure}[h]
\begin{center}
\includegraphics[width=0.65\textwidth]{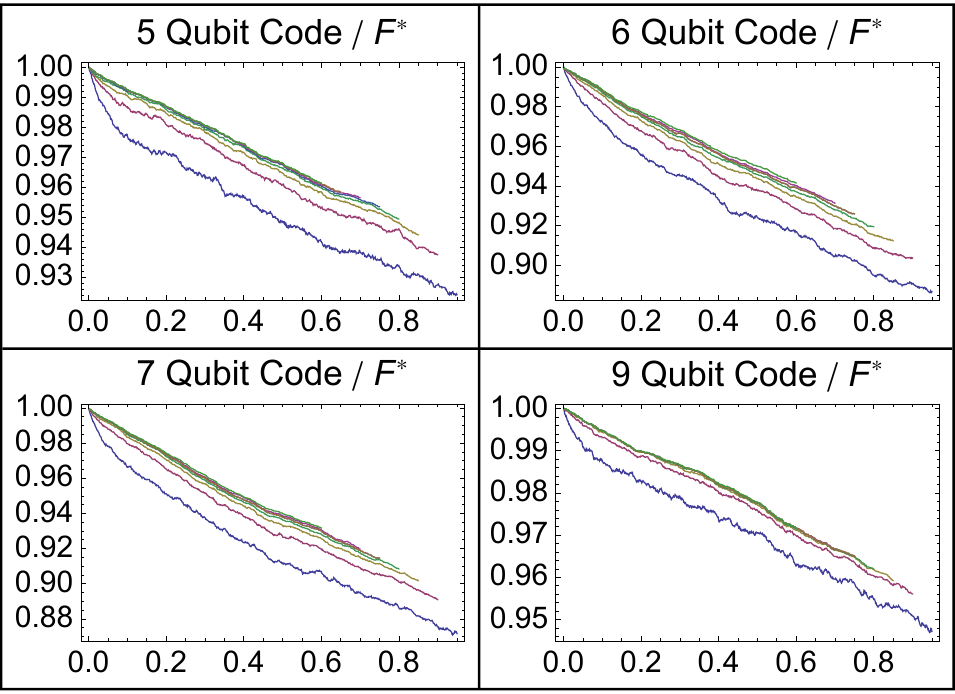}
\caption{\label{fig:fstar-5QB} Averaged finite time-horizon fidelity for $10^4$ quantum trajectories (each) of lossy five-, six-, seven-, and nine-qubit codes with loss parameter $\theta = \frac{\pi}{1000}$ and $\tau = 0.05, 0.1, 0.15, 0.2, 0.25, 0.3, 0.35$ and $0.4$ (bottom to top traces in each sub-panel).}
\end{center}
\end{figure}

\begin{figure}[h]
\begin{center}
\includegraphics[width=0.65\textwidth]{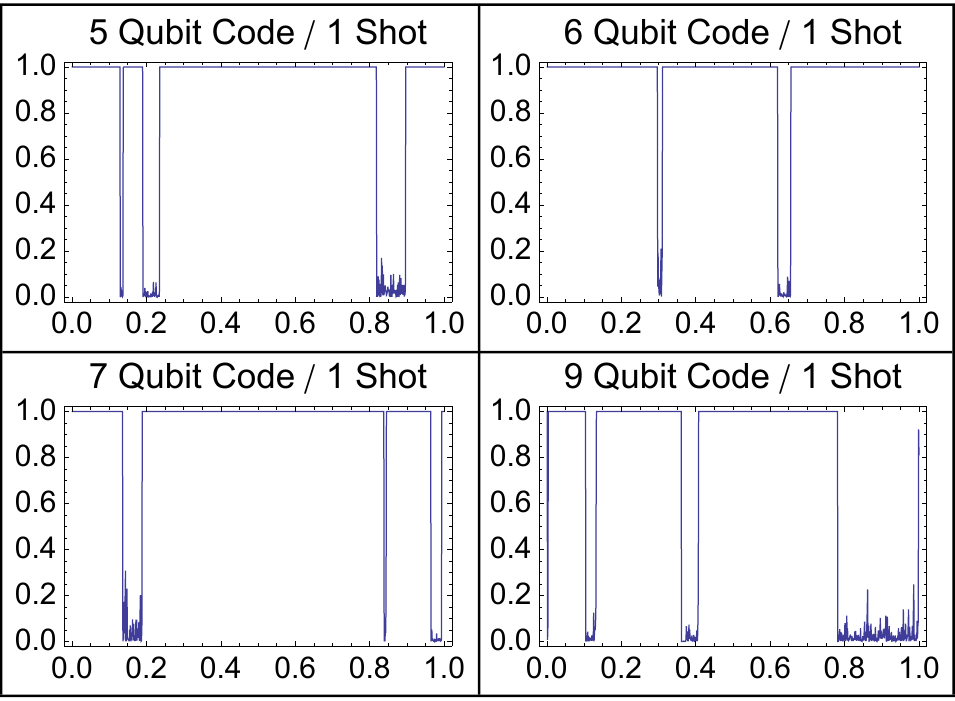}
\caption{\label{fig:single-trajectory}Fidelity trace for single-shot quantum trajectory simulations of five-, six-, seven-, and nine-qubit codes.}
\end{center}
\end{figure}

In Fig.~\ref{fig:lossy-qec} we display the average fidelity decay $F(t)\equiv\langle\Psi_0\vert\rho_t \vert\Psi_0\rangle$ over $10^4$ simulated quantum trajectories (each) for lossy quantum memories implementing the five-, six-, seven- and nine-qubit codes~\cite{Gottesman,Lidar,Bacon2006}, which may be compared directly with analogous results from our prior work on other codes~\cite{Kerc10,Kerc11,Sarma12a}. In Fig.~\ref{fig:fstar-5QB} we display the average decay (again over $10^4$ quantum trajectory simulations for each code) of an alternative performance measure,
\begin{equation*}
F^*_\tau(t) = \mathrm{max}_{t^{*} \in [t, t+\tau]} F(t^*).
\end{equation*}
The quantity $F^*$ represents an easily computable, convenient statistic which we believe is a more meaningful measure of performance of a realistic quantum memory than the canonical fidelity measure. Our definition of $F^*$ is motivated by the observation that, in any realistic quantum memory, there must be a finite latency of error correction. This behavior is clearly illustrated for our class of models by the individual quantum trajectory simulations of fidelity versus time shown in Fig.~\ref{fig:single-trajectory}.

Because of the (random) time delay required for an error-correcting controller with finite-strength measurement and feedback to restore the register state after a decoherence event, fidelity does not decay monotonically along individual quantum trajectories~\cite{Mabu09b}. This behavior has a pronounced effect on the appearance of a simple plot of average $F(t)$ at small $t$ as some trajectories in the ensemble will experience errors at very early times without recovering immediately; this sub-ensemble induces the steep initial decay transient seen for example for $t\lesssim 0.05$ in the nine-qubit panel of Fig.~\ref{fig:lossy-qec}. If we recognize that many (even most) of these trajectories will in fact regain $F\sim 1$ after a finite delay, as seen in the examples of Fig.~\ref{fig:single-trajectory}, it seems intuitive to adopt a performance measure such as $F^*(t)$ that looks ahead over a window of time in each trajectory to check for such a recovery. Of course if an additional error should occur within a given trajectory before the feedback network has had time to correct the initial one, the encoded information can be lost and $F^*(t)\rightarrow 0$ accordingly.

\ack
This work has been supported by the NSF (PHY-1005386), AFOSR (FA9550-11-1-0238) and DARPA-MTO (N66001-11-1-4106).  G.S. would like to thank Dmitri Pavlichin for numerous insightful discussions.

\end{document}